\documentclass[]{aastex631}

\usepackage{bm}
\usepackage{natbib}
\usepackage{mathtools}
\usepackage{amsmath}

\submitjournal{ApJL}

\shorttitle{High-Frequency Oscillation Coronal Heating}
\shortauthors{Lim et al.}

\graphicspath{{./}{figures/}}

\begin{document}
\title{The Role of High-Frequency Transverse Oscillations in Coronal Heating}

\correspondingauthor{Daye Lim}
\email{daye.lim@kuleuven.be}

\author[0000-0001-9914-9080]{Daye Lim}

\affiliation{Centre for mathematical Plasma Astrophysics, Department of Mathematics, KU Leuven, \\
Celestijnenlaan 200B, B-3001 Leuven, Belgium}
\affiliation{Solar-Terrestrial Centre of Excellence – SIDC, Royal Observatory of Belgium, \\
Ringlaan -3- Av. Circulaire, 1180 Brussels, Belgium}

\author[0000-0001-9628-4113]{Tom Van Doorsselaere}
\affiliation{Centre for mathematical Plasma Astrophysics, Department of Mathematics, KU Leuven, \\
Celestijnenlaan 200B, B-3001 Leuven, Belgium}

\author[0000-0003-4052-9462]{David Berghmans}
\affiliation{Solar-Terrestrial Centre of Excellence – SIDC, Royal Observatory of Belgium, \\
Ringlaan -3- Av. Circulaire, 1180 Brussels, Belgium}

\author[0000-0001-5678-9002]{Richard J. Morton}
\affiliation{Department of Maths, Physics and Electrical Engineering, Northumbria University, \\
Newcastle upon Tyne, UK}

\author[0000-0002-6954-2276]{Vaibhav Pant}
\affiliation{Aryabhatta Research Institute of Observational Sciences, Nainital, India}

\author[0000-0002-7762-5629]{Sudip Mandal}
\affiliation{Max Planck Institute for Solar System Research, Justus-von-Liebig-Weg 3, 37077 Göttingen, Germany}
\begin{abstract} 

Transverse oscillations that do not show significant damping in solar coronal loops are found to be ubiquitous. Recently, the discovery of high-frequency transverse oscillations in small-scale loops has been accelerated by the Extreme Ultraviolet Imager onboard Solar Orbiter. We perform a meta-analysis by considering the oscillation parameters reported in the literature. Motivated by the power law of the velocity power spectrum of propagating transverse waves detected with CoMP, we consider the distribution of energy fluxes as a function of oscillation frequencies and the distribution of the number of oscillations as a function of energy fluxes and energies. These distributions are described as a power law. We propose that the power law slope ($\delta=-1.40$) of energy fluxes depending on frequencies could be used for determining whether high-frequency oscillations dominate the total heating ($\delta < 1$) or not ($\delta > 1$). In addition, we found that the oscillation number distribution depending on energy fluxes has a power law slope of $\alpha=1.00$, being less than 2, which means that oscillations with high energy fluxes provide the dominant contribution to the total heating. It is shown that, on average, higher energy fluxes are generated from higher frequency oscillations. The total energy generated by transverse oscillations ranges from about $10^{20}$ to $10^{25}$ erg, corresponding to the energies for nanoflare ($10^{24}-10^{27}$ erg), picoflare ($10^{21}-10^{24}$ erg), and femtoflare ($10^{18}-10^{21}$ erg). The respective slope results imply that high-frequency oscillations could provide the dominant contribution to total coronal heating generated by decayless transverse oscillations.  

\end{abstract}


\section{Introduction}\label{sec:intro}

The coronal heating problem is one of the long-standing astrophysical puzzles \citep{1996SSRv...75..453N, 2015RSPTA.37340256K}. The efforts for solving this have been made based on two mechanisms: magnetic reconnection (DC heating) and waves (AC heating). Regarding DC heating mechanisms, it has been observed that the relationship between the number of energy-release events, i.e., solar flares, by magnetic reconnection and the released energy is described as a power law \citep{2000ApJ...535.1047A}. \citet{1991SoPh..133..357H} first proposed that the power law slope could determine whether flares with lower energies (mainly considered nanoflare in terms of energy) provide the dominant contribution to the total heating generated by flares. Since then, many studies on determining the critical slope have been performed both observationally \citep{1993SoPh..143..275C, 1998A&A...336.1039B, 2000ApJ...529..554P, 2002ESASP.506..501B, 2008ApJ...677..704H, 2016A&A...591A.148J, 2019AstL...45..248U, 2022A&A...661A.149P} and numerically \citep{2013A&A...550A..30B, 2018A&A...617A..50K, 2019ApJ...871..133J}. However, no definite conclusions have been reached regarding the slope in DC heating.

AC heating mechanisms (see \citealt{2020SSRv..216..140V}, for a recent review) have received more attention since it was observationally revealed that transverse magnetohydrodynamic (MHD) oscillations in the solar corona are ubiquitous \citep{2015A&A...583A.136A, 2019NatAs...3..223M}. Transverse oscillations which appear as repetitive transverse displacements of the coronal loop axis have been amply observed with EUV imaging instruments (see \citealt{2021SSRv..217...73N}, for a recent review). Since the commissioning of the Atmospheric Imaging Assembly (AIA; \citealt{2012SoPh..275...17L}) onboard the Solar Dynamics Observatory (SDO; \citealt{2012SoPh..275....3P}), transverse oscillations that do not show significant damping and that have an amplitude less than the minor radius of the loop have been observed \citep{2012ApJ...751L..27W, 2013A&A...552A..57N}. Such persistent oscillations were also observed in Doppler shift \citep{2012ApJ...759..144T} and could last more than 30 oscillation cycles \citep{2022MNRAS.513.1834Z}. This type of oscillation is called decayless oscillation, in comparison to decaying oscillation, which has large amplitudes and rapid damping, excited by an impulsive event \citep{1999ApJ...520..880A, 1999Sci...285..862N, 2015A&A...577A...4Z, 2016A&A...585A.137G, 2019ApJS..241...31N}. The interpretation of observed decayless oscillations as standing modes was suggested by a constant oscillation phase along the loop axis \citep{2013A&A...560A.107A, 2015A&A...583A.136A} and the dependence of the period on the loop length \citep{2015A&A...583A.136A}. \citet{2015A&A...583A.136A} found that decayless oscillations in active regions (ARs) having no explosive events are omnipresent. As an exception, however, there were observations of decayless oscillations that accompany solar flares \citep{2012ApJ...751L..27W, 2013A&A...552A..57N, 2022RAA....22j5017S}. \citet{2021A&A...652L...3M} reported flare-induced decayless oscillations and presented that the flare did not affect the period of oscillations but was likely acting to amplify them.  

Decayless oscillations appeared not only in large-scale loops of a few hundred Mm but also in small-scale loops shorter than a few tens Mm such as coronal bright points \citep{2022ApJ...930...55G}. The discovery of rapid decayless oscillations in smaller loop structures, which had not been observed due to the limitations of AIA's time cadence and spatial resolution, has been accelerated by the unprecedented high-spatial and high-temporal resolution of the High Resolution Imager (HRI) of the Extreme Ultraviolet Imager (EUI; \citealt{2020A&A...642A...8R}) aboard the Solar Orbiter (SolO; \citealt{2020A&A...642A...1M}).  \citet{2023ApJ...946...36P} first discovered high-frequency transverse oscillations with periods of 30 s and 14 s. \citet{2022MNRAS.516.5989Z} presented short-period (1-3 min) decayless oscillations simultaneously observed in both SolO/EUI and SDO/AIA for the first time. As the oscillation results in EUI and AIA are consistent with each other, it confirmed that the observations of high-frequency oscillations by EUI are robust. The expectation that there would be more high-frequency decayless oscillations was proved by \citet{2023ApJ...944....8L} and \citet{2023arXiv230413554S}. \citet{2023ApJ...944....8L} statistically investigated 111 decayless oscillations with periods shorter than 200 s in an AR and they showed a strong dependence of periods on loop lengths. \citet{2023arXiv230413554S} found 42 decayless oscillations with a period range from about 27 s to 276 s in a quiet region and coronal holes. Unlike in \citet{2015A&A...583A.136A} and \citet{2023ApJ...944....8L}, they found that there was no strong correlation between periods and lengths, which was consistent with the lack of correlation found by \citet{2022ApJ...930...55G}.

The omnipresent propagating transverse waves have also been detected by the Doppler velocities of the Coronal Multi-channel Polarimeter (CoMP; \citealt{2008SoPh..247..411T}). \citet{2007Sci...317.1192T} and \citet{2009ApJ...697.1384T} found the power-law relationship in the spectrum of the observed velocity power. It is shown that the power-law index depends on the observed region, e.g., the AR, quiet Sun, or open field region \citep{2016ApJ...828...89M}, however, the power-law index is not influenced by the phase of the solar cycle \citep{2019NatAs...3..223M}. Interestingly, all power spectra showed enhanced power in the frequency region between 3 mHz and 5 mHz \citep{2007Sci...317.1192T, 2009ApJ...697.1384T, 2016ApJ...828...89M, 2019NatAs...3..223M}. This feature proffered the possibility of coronal transverse waves excited by photospheric p-modes \citep{2006ApJ...648L.151J, 2008SoPh..251..251C, 2021ApJ...922..225R, 2023arXiv230401606S}. 

With the new-generation SolO/EUI HRI, we can now observe faster oscillations in smaller loops than in the pre-2020 SDO-era but up till now it was not clear if these high-frequency oscillations are statistically significant. This is important to know as high-resolution, sub-field observations with HRI observations are, in contrast to SDO, sporadic in nature and require dedicated planning of the HRI pointing and imaging cadence. Motivated by the power law relation of propagating transverse waves, we investigate the statistical relationships between energy properties and frequencies of decayless oscillations, including the high-frequency range observed to date. Using our statistical analysis, we propose, to our knowledge for the first time, that the role of high-frequency oscillations in coronal heating could be determined from the power law slope between energy fluxes and oscillation frequencies. It could be considered a counterpart of similar statistics arguments in the nanoflare heating theory \citep{1991SoPh..133..357H}. The relationships between the occurrence number of oscillations and their energy properties are also presented. 

\section{Energy Property Distributions}\label{sec:energyproperties}

We perform a meta-analysis of decayless oscillations reported in the literature. We only selected decayless oscillations of coronal loops, where no solar flare greater than C1.0 occurred, from the literature that presented periods, displacement amplitudes, and loop lengths. The literature with details is summarized in Table \ref{tab:list}. We need to note that among the oscillations we consider, while there are cases where the decayless oscillations were clearly identified as standing modes, there are cases where the interpretation of whether observed oscillations are standing or propagating waves still remains open \citep{2022ApJ...930...55G, 2023ApJ...946...36P, 2023arXiv230413554S}. In this study, all decayless oscillations we consider in this study are assumed to be standing modes. Our data samples include 120 oscillations observed by SDO/AIA 171 \AA\;and 170 oscillations by the SolO/EUI HRI 174 \AA. The data consist of 79 oscillations in the quiet Sun and coronal holes and 211 oscillations in ARs, showing a bias to AR loop oscillations. We assume that standing decayless oscillations are omnipresent regardless of their periods and where they occur.

\begin{table}[h]
\centering
\caption{The decayless transverse oscillations reported in the literature that were considered for the analysis. The first column shows the reference. The instrument, time cadence, and pixel scale of the observation are mentioned in the second, third, and fourth columns respectively. The fifth and sixth columns show the number of oscillation events considered in this study and the region where the oscillating coronal loops are embedded. Note that \citet{2018ApJ...854L...5D} found the second harmonic of decayless oscillation in the solar corona, the first harmonic of decayless oscillation was only considered here. \citet{2022MNRAS.516.5989Z} reported oscillations detected with SDO/AIA and SolO/EUI simultaneously, but we only used oscillation results observed by EUI. \citet{2023ApJ...946...36P} presented two possible loop lengths for each oscillation, thus we considered the average value of loop lengths. \citet{2023ApJ...944....8L} reported a total number of 111 decayless oscillations, however, we only consider oscillations with periods longer than 20 s considering the temporal cadence of 3 s of the observation that they considered. In the case of literature using multiple observations with different time cadences and pixel scales, the minimum values are listed.}
\label{tab:list}
\begin{tabular}{cccccc}
\hline
Reference & Instrument & Cadence (s) & Pixel scale (km) & \# of oscillations & Region\\ 
\hline
{\citet{2013A\string&A...560A.107A}} & AIA & 12 & 435 &  10 & Active region              \\
{\citet{2015A\string&A...583A.136A}} & AIA & 12 & 435 &  72 & Active region              \\
{\citet{2018ApJ...854L...5D}}        & AIA & 12 & 435 &   1 & Quiet Sun                  \\
{\citet{2022ApJ...930...55G}}        & AIA & 12 & 435 &  31 & Quiet Sun                  \\
{\citet{2022MNRAS.513.1834Z}}        & AIA & 12 & 435 &   6 & Quiet Sun \& active region \\
{\citet{2023ApJ...946...36P}}        & EUI &  2 & 200 &   2 & Quiet Sun                  \\
{\citet{2022MNRAS.516.5989Z}}        & EUI &  5 & 306 &   7 & Active region              \\
{\citet{2022A\string&A...666L...2M}} & EUI &  5 & 185 &  15 & Active region              \\
{\citet{2023ApJ...944....8L}}        & EUI &  3 & 135 & 104 & Active region              \\
{\citet{2023arXiv230413554S}}        & EUI &  3 & 119 &  42 & Quiet Sun \& coronal hole  \\ 
\hline
\end{tabular}
\end{table}

For analysing the energy properties of transverse oscillations observed with imaging instruments, we adopt Equation (11) for the energy ($E$ with the physical unit of J) and Equation (19) for the energy flux ($F$ with the physical unit of $\text{W}\,\text{m}^{-2}$) of kink oscillations from \citet{2014ApJ...795...18V}. By assuming that the oscillating loops are much denser than the surroundings and for the kink oscillations the phase speed is equal to the group speed, the equations represent the lower limit of the total energy and energy flux, which are as follows
\begin{equation}\label{eq:flux}
    E=(\pi R^{2}L)\frac{1}{2}\rho \left(\frac{2\pi A}{P}\right)^{2},
\end{equation}
\begin{equation}\label{eq:energy}
    F=\frac{1}{2}f\rho \left(\frac{2\pi A}{P}\right)^{2}\left(\frac{2L}{P}\right),
\end{equation}
where $R$, $L$, and $\rho$ are the minor radius, length, and plasma density of the oscillating loop respectively, $f$ is the filling factor, and $A$ and $P$ are the displacement amplitude and period of the oscillation. The latter are the result of fitting the observations with a harmonic function. It has long been argued that coronal loops are composed of bundles of fine-scale flux tubes \citep{2014LRSP...11....4R, 2016ApJ...823...82M}. Since the transverse wave energy is localized to the flux tubes, one should take fine structuring into account for calculating energy properties. The two Equations \label{eq:flux} and \label{eq:energy} are only valid for small filling factors up to 10\% as described in \citet{2014ApJ...795...18V}. Observationally, at a temperature of about 1 MK which is corresponding to the temperature of AIA 171 \AA\;and HRI 174 \AA\;channels, it was found that filling factors are around 1.5\% for coronal bright points \citep{2008A&A...491..561D} and 10\% for AR loops \citep{2008ApJ...686L.131W}. In order to satisfy the theoretical and observational validities, we chose the upper limit of the filling factor of 10\% for all loops considered in this study.

For calculating these two quantities, we use the analysed parameters, $A$, $P$, and $L$, presented in the papers listed in Table \ref{tab:list}. The lengths of the oscillating loops range from a few Mm to a few hundred Mm. In the solar corona, estimating the density is not trivial \citep{2009ApJ...700..762W, 2018LRSP...15....5D}. We consider density stratification by assuming that all oscillating loops have the same uniform temperature and gravitational acceleration. The oscillating loops were observed in channels representing similar temperatures and were not associated with any solar event. The major radius of coronal loops is much smaller than the radius of the Sun. Thus, the assumptions can be considered reasonable. For simplicity, we assume that coronal loops have a semi-circle shape and the loop's plane is perpendicular to the solar surface. Then the height ($h$) of the loop apex is equal to the major radius ($R_{\text{L}}=L/\pi$) of the loop. The density ($\rho$) at the loop apex is as follows,
\begin{equation}\label{eq:density}
\rho=\rho_{\text{B}}\exp{\left(-\frac{R_{\text{L}}}{H}\right)},
\end{equation}
where $\rho_{\text{B}} = 3.3\times10^{-12}\,\text{kg}\,\text{m}^{-3}$ is a plasma density at the footpoints of 1 MK loops \citep{2003A&A...406L...5D}, which is assumed to be the same for all loops, and $H$ the typical hydrostatic pressure scale height (about 47 Mm) at 1 MK. We assumed that the average density of the loop is equal to the density ($\rho$) at the apex of the loop. As a result, the ratio of the derived average density in the longest loop (about 600 Mm) to that in the smallest loop (about 3 Mm) is around 40. This ratio is consistent with the density ratio of typical giant arches (100 - 1000 Mm) to typical bright points (1 - 10 Mm) defined in X-ray coronal loops \citep{2014LRSP...11....4R}. 
Unfortunately, most literature did not include the minor radius of the oscillating loops. Equations \ref{eq:flux} and \ref{eq:energy} were obtained by using the long-wavelength limit \citep{2013ApJ...768..191G, 2013ApJ...771...74G, 2014ApJ...795...18V}. Thus, we assume that the minor radius, $R=0.5$ Mm, is the same for all oscillating loops in order to satisfy the thin flux tube approximation for various loop lengths.

\begin{figure}
\plotone{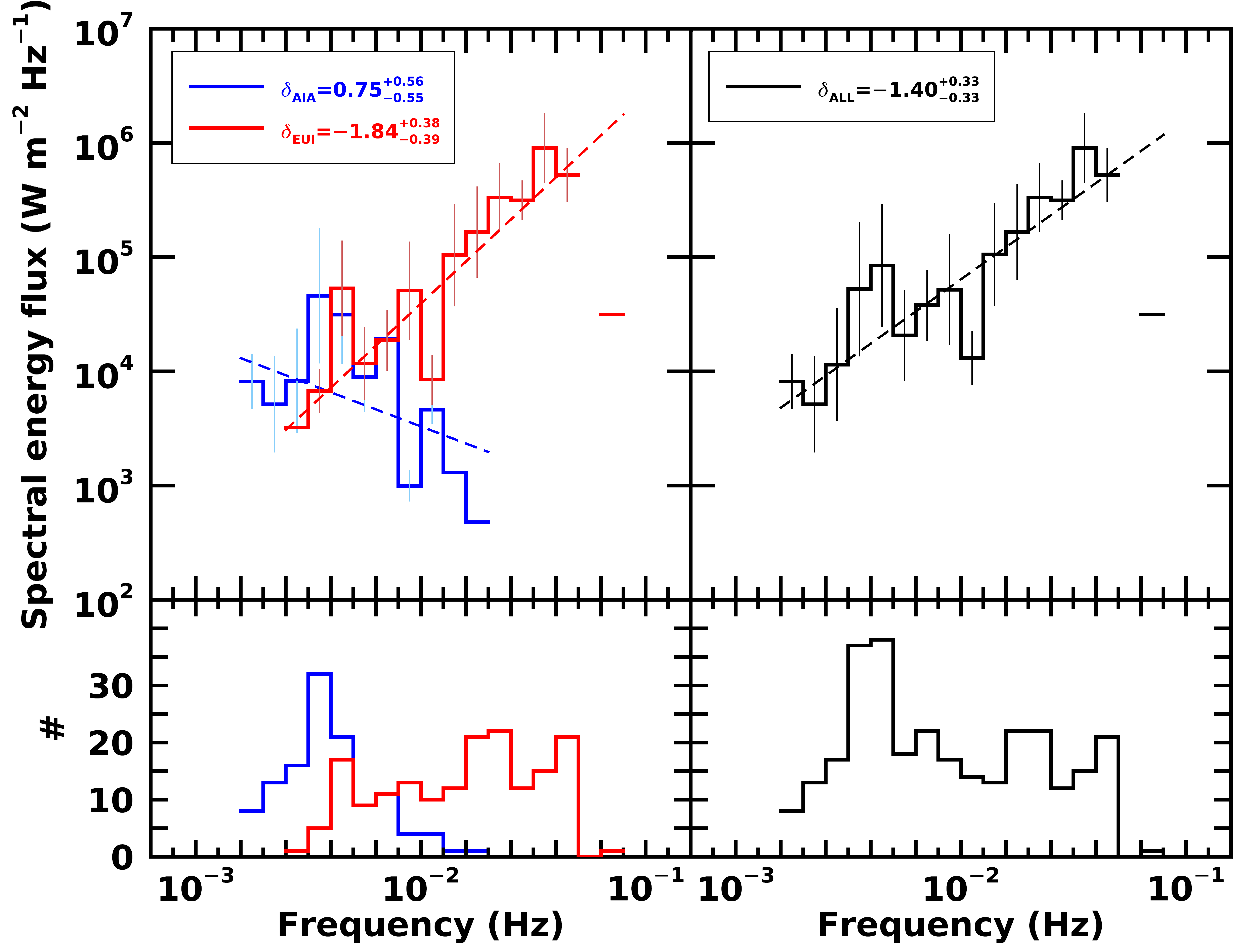}
\caption{The distribution, $s(\omega)$, of spectral energy fluxes as a function of oscillation frequencies (top panels) and the number of oscillations for each frequency bin (bottom panels). The vertical bars show an uncertainty $\sigma_{\omega}=\sigma_{F_{i}}\sqrt{n_{\omega_{i}}}$). A bin size of $\overline{\omega}=0.1$ has been considered. The best fits of distributions are shown in dashed lines. The fitting was only considered for bins with a number greater than 1 of cases. Blue, red, and black colors represent observations from AIA, EUI, and both of them, respectively. The power-law slopes are $\delta_{\text{AIA}}=0.75^{+0.56}_{-0.55}$, $\delta_{\text{EUI}}=-1.84^{+0.38}_{-0.39}$, and $\delta_{\text{ALL}}=-1.40^{+0.33}_{-0.33}$, respectively.}
\label{fig:frequency}
\end{figure}

For investigating the relationship between the energy flux and frequency, the logarithm of the energy flux of each oscillation is considered. Then they have been binned with a constant bin size ($\overline{\omega}=0.1$) of oscillation frequencies in log scale. The estimated total spectral energy flux in each bin, $s(\omega)$, is given by
\begin{equation}\label{eq:spectralfulx}
s(\omega)=\frac{\displaystyle\sum_{k=1}^{n_{\omega_{i}}} F_{i,\,k}}{\omega_{i+\frac{1}{2}}-\omega_{i-\frac{1}{2}}},
\end{equation}
where $\omega_{i+\frac{1}{2}}=\omega_{i}+10^{\overline{\omega}/2}$ and $\omega_{i-\frac{1}{2}}=\omega_{i}-10^{\overline{\omega}/2}$ are the maximum and minimum frequency values of $i$-th bin respectively. $n_{\omega_{i}}$ and $F_{i,\,k}$ are the number of oscillations and the energy flux of each oscillation in the $i$-th bin, respectively.
Figure \ref{fig:frequency} shows the distribution of spectral energy fluxes as a function of oscillation frequencies. The distribution has an uncertainty according to the standard error for the sum of the number of events, which corresponds the standard deviation ($\sigma_{F_{i}}$) of energy fluxes times the square root of the number of oscillations per bin. In order to estimate the best power law fit ($s\propto\omega^{-\delta}$) of the distribution and its credible interval, we use the Solar Bayesian Analysis Toolkit (SoBAT; \citealt{2021ApJS..252...11A}). The logarithmic uncertainties of each bin are taken into account in the fit. The fitting was only considered for bins with a number greater than 1 in cases. It is shown that the power-law slope ($\delta$) of spectral energy flux from transverse oscillations observed by AIA is around $\delta_{\text{AIA}}=0.75^{+0.56}_{-0.55}$ between frequency bins of about 0.002 and 0.02 Hz. In the case of EUI oscillations with frequencies ranging from about 0.003 to 0.07 Hz, on the contrary, the slope of spectral energy flux is around $\delta_{\text{EUI}}=-1.84^{+0.38}_{-0.39}$. When we consider all observed transverse oscillations, a general power-law form has a slope of $\delta_{\text{ALL}}=-1.40^{+0.33}_{-0.33}$ between the frequency of about 0.002 and 0.07 Hz.

\begin{figure}
\centering
\includegraphics[width=0.6\textwidth,clip=]{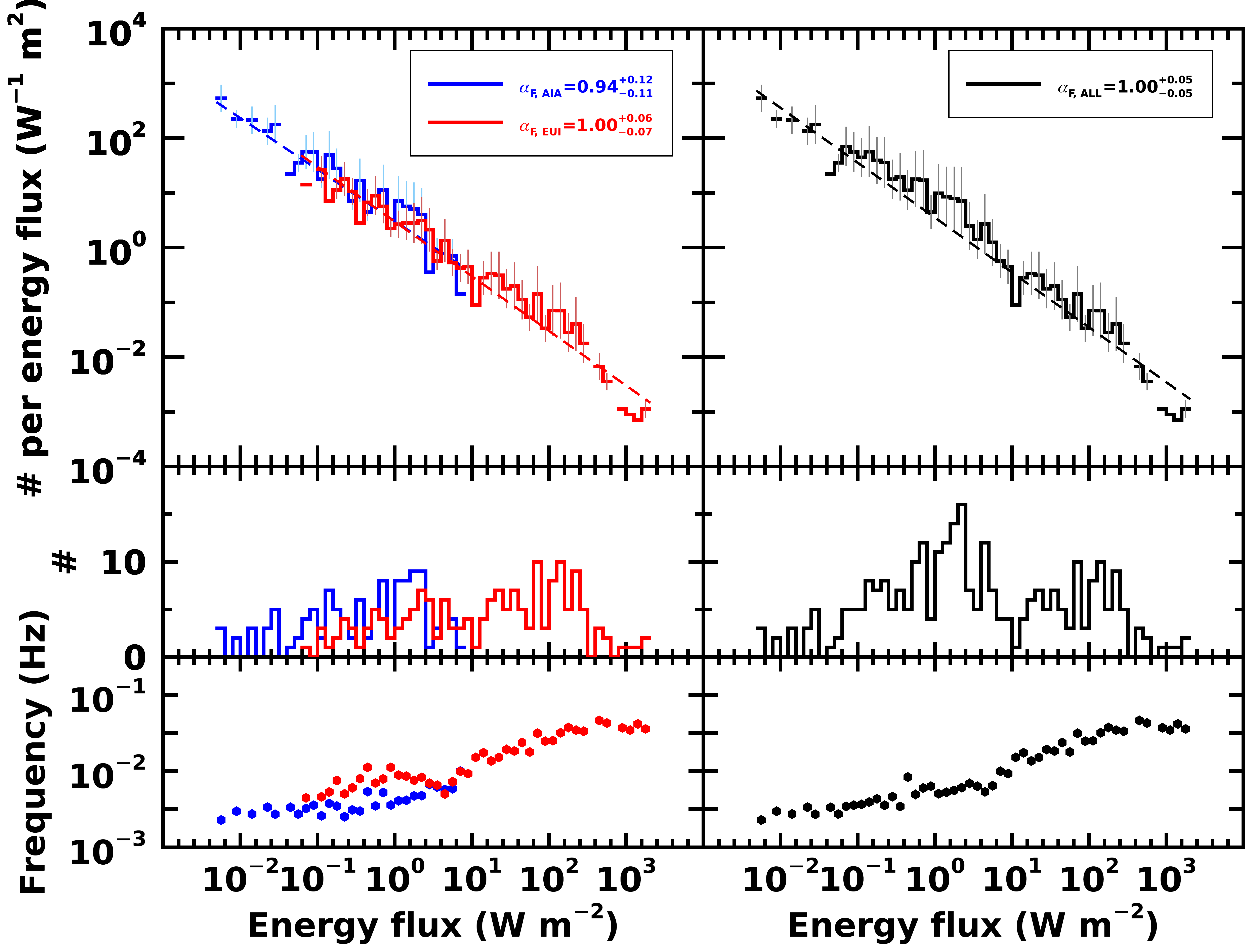}
\includegraphics[width=0.6\textwidth,clip=]{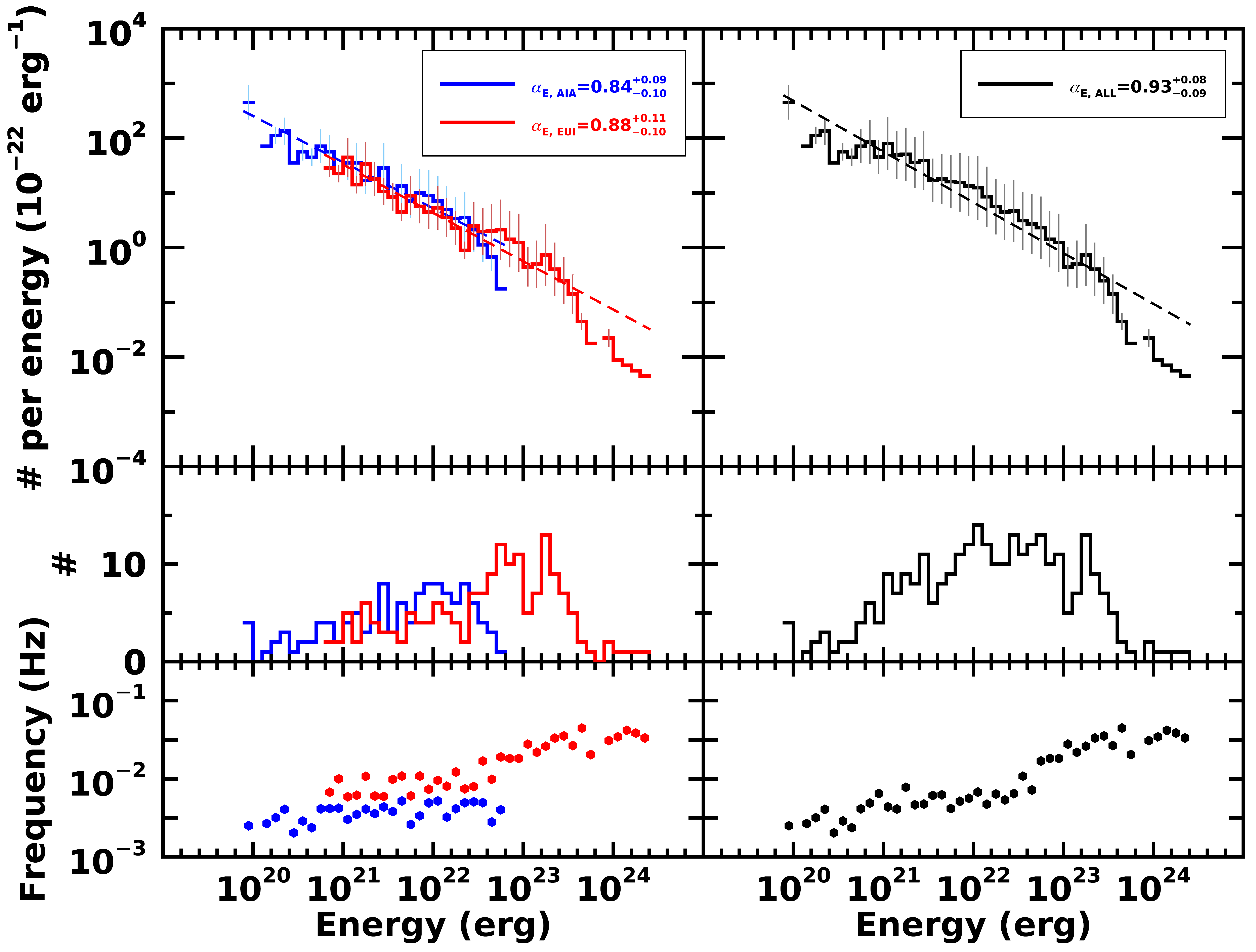}
\caption{Top: the top panels show the distribution of oscillation number per energy flux, $n(F)$, and its power-law fitting (dashed lines). The fitting was only considered for bins with a non-zero amount of cases. A bin size of $\overline{F}=0.1$ for the energy flux distribution has been considered. Blue, red, and black colors represent observations from AIA, EUI, and both of them, respectively. The power law slopes are $\alpha_{F, \text{AIA}}=0.94^{+0.12}_{-0.11}$, $\alpha_{F, \text{EUI}}=1.00^{+0.06}_{-0.07}$, and $\alpha_{F, \text{ALL}}=1.00^{+0.05}_{-0.05}$. The middle and bottom panels show the number of oscillations and mean oscillation frequencies for each energy flux bin. The vertical bars show uncertainties $\sigma_{n(F)}=\sqrt{n_{F_{j}}}$. Bottom: the same as in the top panel, but for energies, $n(E)$, with uncertainties $\sigma_{n(E)}=\sqrt{n_{E_{l}}}$. A bin size of $\overline{E}=0.1$ for the energy distribution has been considered. The power law slopes are $\alpha_{E, \text{AIA}}=0.84^{+0.09}_{-0.10}$, $\alpha_{E, \text{EUI}}=0.88^{+0.11}_{-0.10}$, and $\alpha_{E, \text{ALL}}=0.93^{+0.08}_{-0.09}$.}
\label{fig:energy}
\end{figure}

Additionally, we consider the occurrence number of oscillations as a function of the energy flux, $n(F)$ (with the physical units of $\text{W}^{-1}\,\text{m}^{2}$), and energy, $n(E)$ (with the physical units of $\text{erg}^{-1}$), respectively, which are similar statistics in solar flare distributions. The logarithm of the energy flux and energy have been binned with a constant bin size ($\overline{F}$ and $\overline{E}=0.1$). The estimated oscillation numbers in each energy flux bin and energy bin are as follows
\begin{equation}\label{eq:occurrence_energyflux}
n(F)=\frac{n_{F_{j}}}{F_{j}},
\end{equation}
\begin{equation}\label{eq:occurrence_energy}
n(E)=\frac{n_{E_{l}}}{E_{l}},
\end{equation}
where $F_{j}$ and $E_{l}$ are the center energy flux and energy value of $j$-th and $l$-th bin respectively, $n_{F_{j}}$ is the number of oscillations in an energy flux range $F_{j}-\overline{F}/2 \leq F < F_{j}+\overline{F}/2$, and $n_{E_{l}}$ is the number of oscillations in an energy range $E_{l}-\overline{E}/2 \leq E < E_{l}+\overline{E}/2$. As a result, the distributions of oscillation numbers as a function of energy fluxes and energies are shown in Figure \ref{fig:energy}. Each distribution has uncertainties according to the square root of the number of each bin \citep{2002ApJ...572.1048A}. The fitting with power laws ($n(F)\propto F^{-\alpha_{F}}$ and $n(E)\propto E^{-\alpha_{E}}$) was only considered for bins with a number greater than 1 in cases and with taking into account the logarithmic uncertainties. It is shown that the observed standing transverse oscillations to date have an energy flux ranging from about 0.01 to 1780 $\text{W}\,\text{m}^{-2}$ and an energy ranging from about $10^{20}$ to $10^{25}$ erg. This energy range corresponds to femtoflare ($10^{18}-10^{21}$ erg), picoflare ($10^{21}-10^{24}$ erg), and nanoflare ($10^{24}-10^{27}$ erg). We found that the power-law slopes of number per energy flux distributions are $\alpha_{F, \text{AIA}}=0.94^{+0.12}_{-0.11}$ for AIA oscillations, $\alpha_{F, \text{EUI}}=1.00^{+0.06}_{-0.07}$ for EUI oscillations, and $\alpha_{F, \text{ALL}}=1.00^{+0.05}_{-0.05}$ for all oscillations, respectively. In the case of number per energy, the power law slopes are $\alpha_{E, \text{AIA}}=0.84^{+0.09}_{-0.10}$ for AIA, $\alpha_{E, \text{EUI}}=0.88^{+0.11}_{-0.10}$ for EUI, and $\alpha_{E, \text{ALL}}=0.93^{+0.08}_{-0.09}$ for all observations. We found that the higher energy fluxes and higher energies are generated by higher frequency oscillations on average. 

\section{High-Frequency oscillation Heating Theory} \label{sec:ACheating}

If the energy flux is generated by transverse oscillations with frequencies ranging from $\omega_{\text{min}}$ to $\omega_{\text{max}}$, then the total energy flux, $F_{\omega}\,(\text{W}\,\text{m}^{-2})$, is equal to the integral of oscillation frequencies as follows
\begin{equation}\label{eq:totalenergyflux_freq_1}
F_{\omega}=\int_{\omega_{\text{min}}}^{\omega_{\text{max}}}s(\omega)d\omega,
\end{equation}
where $\omega$ is oscillation frequency (Hz) and $s(\omega)$ is spectral energy flux which is the energy flux per frequency ($\text{W}\,\text{m}^{-2}\,\text{Hz}^{-1}$). If $s(\omega)$ is described as a form of a power law as follows
\begin{equation}\label{eq:spectralflux_powerlaw}
s(\omega)=s_{0}\omega^{-\delta},
\end{equation}
where $s_{0}$ is a scaling constant and $\delta$ is a power-law slope, then, 
\begin{equation}\label{eq:totalenergyflux_freq_2}
F_{\omega}=\int_{\omega_{\text{min}}}^{\omega_{\text{max}}}s_{0}\omega^{-\delta}d\omega=\left.\frac{s_{0}}{-\delta+1}\omega^{-\delta+1}\right\vert_{\omega_{\text{min}}}^{\omega_{\text{max}}}.
\end{equation}
Thus, the total energy flux becomes
\begin{equation}\label{eq:totalenergyflux_freq_3}
F_{\omega}=\frac{s_{0}}{-\delta+1}(\omega_{\text{max}}^{-\delta+1}-\omega_{\text{min}}^{-\delta+1}).
\end{equation}
If $\delta < 1$, 
\begin{equation}\label{eq:totalenergyflux_freq_4}
F_{\omega}\approx\frac{s_{0}}{-\delta+1}\omega_{\text{max}}^{-\delta+1},
\end{equation}
implying that oscillations with the frequency $\omega_{\text{max}}$ provide the dominant heating contribution to $F_{\omega}$, whereas if $\delta > 1$, 
\begin{equation}\label{eq:totalenergyflux_freq_5}
F_{\omega}\approx\frac{s_{0}}{\delta-1}\omega_{\text{min}}^{-\delta+1},
\end{equation}
indicating that oscillations with the frequency $\omega_{\text{min}}$ dominate the heating $F_{\omega}$.

Based on this formula, the slope results shown in Figure \ref{fig:frequency} could be interpreted as follows. If we assume that the total energy flux from all observed transverse oscillations is $F_{\omega,\text{ALL}}$, the slope of $\delta_{\text{ALL}}=-1.40^{+0.33}_{-0.33}$, which is less than the critical slope of 1, between frequencies 0.002 and 0.07 Hz suggests that the transverse oscillations with the frequency of 0.07 Hz give the dominant contribution to the total heating $F_{\omega,\text{ALL}}$ compared to the contribution of oscillations with the frequency of 0.002 Hz. It implies that higher-frequency transverse oscillations could play a key role in contributing to total wave-based heating. If limited to the frequency range of the oscillations observed by AIA and EUI respectively, the slopes of $\delta_{\text{AIA}}=0.75^{+0.56}_{-0.55}$ and $\delta_{\text{EUI}}=-1.84^{+0.38}_{-0.39}$ are less than 1, indicating that high-frequency oscillations mainly contribute to total heating from AIA and EUI oscillations respectively.

If transverse oscillations can generate energy fluxes ranging from $F_{\text{min}}$ to $F_{\text{max}}$, then the total energy flux ($F_{n}$) from transverse oscillations could be equal to the integral of oscillation energy fluxes as follows
\begin{equation}\label{eq:totalenergyflux_flux_1}
F_{n}=\int_{F_{\text{min}}}^{F_{\text{max}}}n(F)FdF,
\end{equation}
where $F$ is the energy flux generated by the oscillation and $n(F)$ is the number of oscillations per energy flux ($\text{W}^{-1}\,\text{m}^{2}$). If $n(F)$ follows a power law, which is given by
\begin{equation}\label{eq:number_powerlaw}
n(F)=n_{F0}F^{-\alpha_{F}},
\end{equation}
where $n_{F0}$ is a scaling constant and $\alpha_{F}$ is a power-law slope, then, 
\begin{equation}\label{eq:totalenergyflux_flux_2}
F_{n}=\frac{n_{F0}}{-\alpha_{F}+2}(F_{\text{max}}^{-\alpha_{F}+2}-F_{\text{min}}^{-\alpha_{F}+2}).
\end{equation}
If $\alpha_{F} < 2$, 
\begin{equation}\label{eq:totalenergyflux_flux_3}
F_{n}\approx\frac{n_{F0}}{-\alpha_{F}+2}F_{\text{max}}^{-\alpha_{F}+2},
\end{equation}
implying that oscillations with the energy flux of $F_{\text{max}}$ provide the dominant heating contribution to $F_{n}$, whereas if $\alpha_{F} > 2$, 
\begin{equation}\label{eq:totalenergyflux_flux_4}
F_{n}\approx\frac{n_{F0}}{\alpha_{F}-2}F_{\text{min}}^{-\alpha_{F}+2},
\end{equation}
suggesting that oscillations with the energy flux of $F_{\text{min}}$ dominate the heating $F_{n}$. This relationship could be applied to the total energy in the same way.

It is found that all slopes from the occurrence number-energy flux relation and the occurrence number-energy relation are less than the critical value of 2. Based on the formula, these results could be interpreted that oscillations with higher energy flux and oscillations with higher energy provide the relatively dominant contribution to the total heating $F_{n}$ and $E_{n}$ respectively compared to oscillations with lower energy fluxes and energies. As mentioned above, higher energy fluxes and energies are generated by higher frequency oscillation as shown in Figure \ref{fig:energy}. Consequently, both approaches that the energy flux-frequency relation and occurrence number-energy property relation indicate that high-frequency oscillations give the most important significant contribution to coronal heating. The interpretation of these results from the perspective of nanoflare heating theory is consistent with the interpretation of the results from the slope of energy flux and frequency distribution that is proposed in this study. 

Since the total energy flux ($F_{\omega}$ and $F_{n}$) generated by transverse oscillations depends on the constant term ($s_{0}$ and $n_{F0}$), the slope result does not immediately lead to the conclusion that high-frequency oscillations could be enough to heat the solar corona. Using the empirical fitting results for all oscillations, $s_{0}\approx4.07^{+14.13}_{-3.18}\times10^{7}$ and $n_{F0}\approx3.55^{+0.82}_{-0.67}$, the observed highest frequency ($\omega_{\text{max}}\approx0.07$ Hz), and the derived highest energy flux ($F_{\text{max}}\approx1780$ $\text{W}\,\text{m}^{-2}$), we estimate the total energy flux of about $2.87^{+32.89}_{-2.64}\times10^{4}$ $\text{W}\,\text{m}^{-2}$ from Equation \ref{eq:totalenergyflux_freq_4} and about $6.32^{+4.44}_{-2.60}\times10^{3}$ $\text{W}\,\text{m}^{-2}$ from Equation \ref{eq:totalenergyflux_flux_3} respectively. The total energy flux values estimated from each power law model are comparable to the sum of the energy fluxes ($1.85\times10^{4}$ $\text{W}\,\text{m}^{-2}$) of each oscillation. Given the energy losses of roughly $10^{4}$ $\text{W}\,\text{m}^{-2}$ in ARs, 300 $\text{W}\,\text{m}^{-2}$ in the quiet Sun, and  800 $\text{W}\,\text{m}^{-2}$ in coronal holes \citep{1977ARA&A..15..363W, 2006SoPh..234...41K}, both values indicate that the quiet Sun and coronal holes could be sufficiently heated by high frequency standing transverse oscillation, and these values indicate that these could also be sufficient heating for ARs. It should be kept in mind that the power law slopes and the estimates of the total energy flux could be influenced by the limited data sample and the assumptions used in this study.

\section{Discussion and Conclusions}\label{sec:discussion}

We performed a statistical analysis of decayless oscillations observed by the SDO/AIA and SolO/EUI in the literature. The frequency of the oscillations ranges from about 0.002 to 0.07 Hz. The transverse oscillations generated energy fluxes ranging from about 0.01 to 1780 $\text{W}\,\text{m}^{-2}$ and energies ranging from about $10^{20}$ to $10^{25}$ erg. 
The relationship between the spectral energy flux and the frequency has a power law with the slope of $\delta_{\text{AIA}}=0.75^{+0.56}_{-0.55}$ for AIA oscillations, $\delta_{\text{EUI}}=-1.84^{+0.38}_{-0.39}$ for EUI oscillations, and $\delta_{\text{ALL}}=-1.40^{+0.33}_{-0.33}$ for all oscillations. Based on the critical slope of 1, which we proposed to determine which of low- and high-frequency oscillations contributes predominantly to heating, it was shown that high-frequency oscillations could provide the dominant contribution to the total heating generated by decayless oscillations. 
The distributions of the occurrence number of oscillations as a function of the energy flux and energy could be also described as the power law with slopes less than 2 for all cases, which means that oscillations with high energy flux and high energy dominate the total heating from oscillations. It was found that high-frequency oscillations generate high energy flux and energy. The interpreted result based on the nanoflare heating theory supported the vital role of high-frequency oscillations revealed from the slope result of the energy flux-frequency relation proposed in this study.

We assumed the filling factor to be 10\% for all oscillating loops when we calculated the energy flux and energy of each oscillation. Even if the filling factor is considered as 100\% the energy fluxes of all oscillations are increased overall, and the slope result will not change, only the total energy flux value will increase.

It seemed that the power law tendency of AIA and EUI oscillations in spectral energy fluxes depending on frequencies is opposite, as shown in Figure \ref{fig:frequency}. Since at least five data points are required to measure the oscillation period from the observation, the observable period by AIA with a temporal cadence of 12 s should be longer than 1 minute. It is noted that most of the oscillations observed with EUI were from studies focusing on high-frequency oscillations in small-scale loops that could not be observed with AIA. Thus, the discrepancy between the two tendencies may be due to the cutoff frequencies of AIA and EUI respectively and this would be resolved if there are more detected oscillations in the future. 

Our results showed that high-frequency transverse oscillations could play a key role in coronal heating compared to low-frequency oscillations. The total energy flux generated by decayless oscillations was sufficient to heat the quiet Sun and comparable to the heating requirements of ARs. The number of higher frequency transverse oscillations than 0.07 Hz observed to date is extremely sparse. It could be expected to observe coronal transverse oscillations in the higher frequency range from missions such as the future high-cadence (i.e., less than 3 s) HRI campaigns of the SolO and the Multi-slit Solar Explorer with high temporal resolution (down to ~0.5 s; \citealt{2022ApJ...926...52D}). 

\citet{2016ApJ...828...89M} estimated the spectra with a frequency range between 0.0002 and 0.02 Hz for energy flux of propagating transverse waves in ARs, quiet Sun, and open field regions. The estimated slope for open field regions that are not considered in the meta-analysis is roughly 1.2, which is larger than the critical slope ($\delta=1$) indicating the relatively dominant contribution of lower frequency waves in coronal heating. Note that the slopes of the spectra for energy flux were not presented in \citet{2016ApJ...828...89M}, we estimated the slope ourselves roughly. They also showed that the slopes of spectra for energy flux varied between different coronal regions. Thus, our slope results obtained by considering the global coronal region may differ when considering each coronal region separately. More observed decayless transverse oscillations would allow a discussion in detail of our results, along with results separated by coronal region, in the near future.\\

We are grateful to Hugh Hudson and the referee for constructive comments. We thank the NASA’s Living With a Star Program, which SDO is part of, with the AIA instrument on-board. Solar Orbiter is a space mission of international collaboration between ESA and NASA, operated by ESA. The EUI instrument was built by CSL, IAS, MPS, MSSL\/UCL, PMOD\/WRC, ROB, LCF\/IO with funding from the Belgian Federal Science Policy Office (BELSPO/PRODEX PEA C4000134088); the Centre National d’Etudes Spatiales (CNES); the UK Space Agency (UKSA); the Bundesministerium für Wirtschaft und Energie (BMWi) through the Deutsches Zentrum für Luft- und Raumfahrt (DLR); and the Swiss Space Office (SSO). TVD was supported by the European Research Council (ERC) under the European Union's Horizon 2020 research and innovation programme (grant agreement No 724326), the C1 grant TRACEspace of Internal Funds KU Leuven, and a Senior Research Project (G088021N) of the FWO Vlaanderen. The research benefitted greatly from discussions at ISSI. DL was supported by a Senior Research Project (G088021N) of the FWO Vlaanderen. VP was supported by SERB start-up research grant (File no. SRG/2022/001687). RJM is supported by a UKRI Future Leader Fellowship (RiPSAW—MR/T019891/1).

\vspace{5mm}
\facilities{SDO, SolO}

\software{SolarSoftWare \citep{1998SoPh..182..497F} 
          , SoBAT \citep{2021ApJS..252...11A}
          }

\appendix
\section{Power law fitting}

\begin{figure}
\centering
\includegraphics[width=\textwidth,clip=]{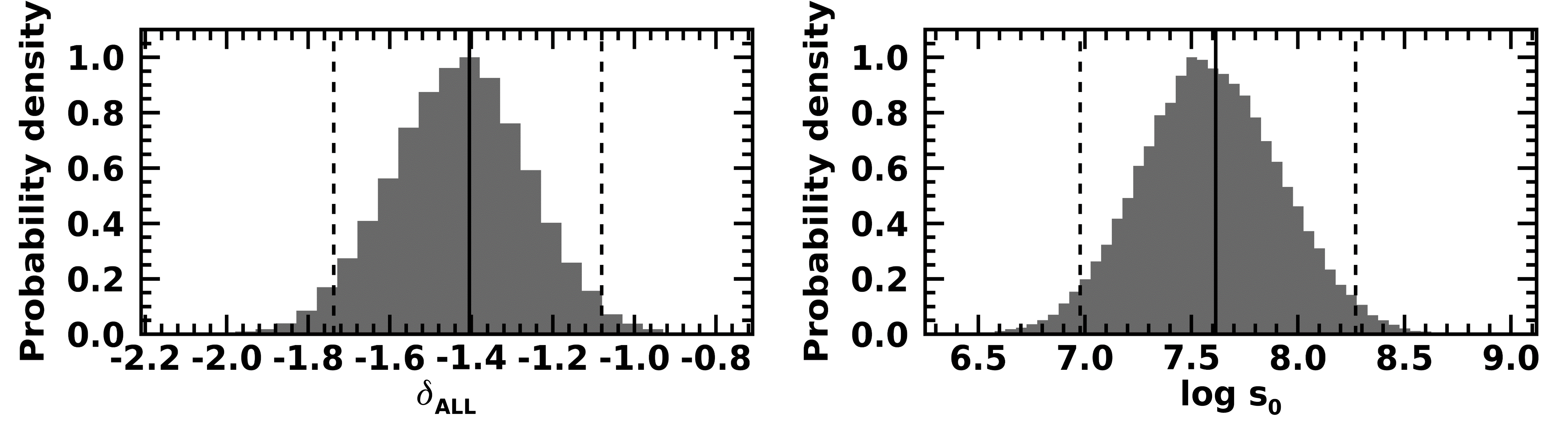}
\includegraphics[width=\textwidth,clip=]{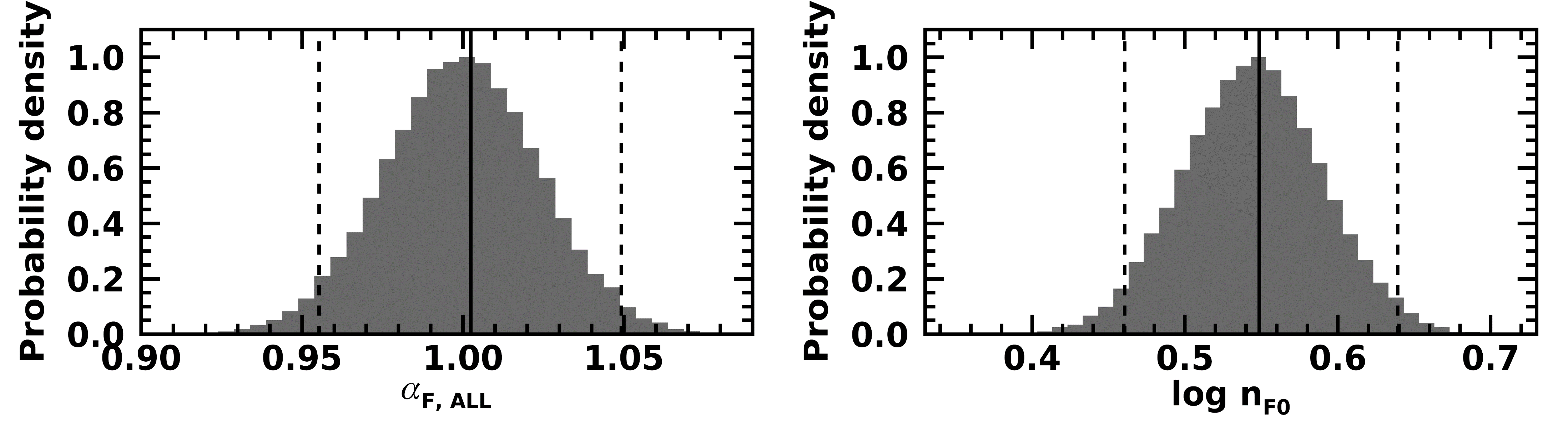}
\includegraphics[width=\textwidth,clip=]{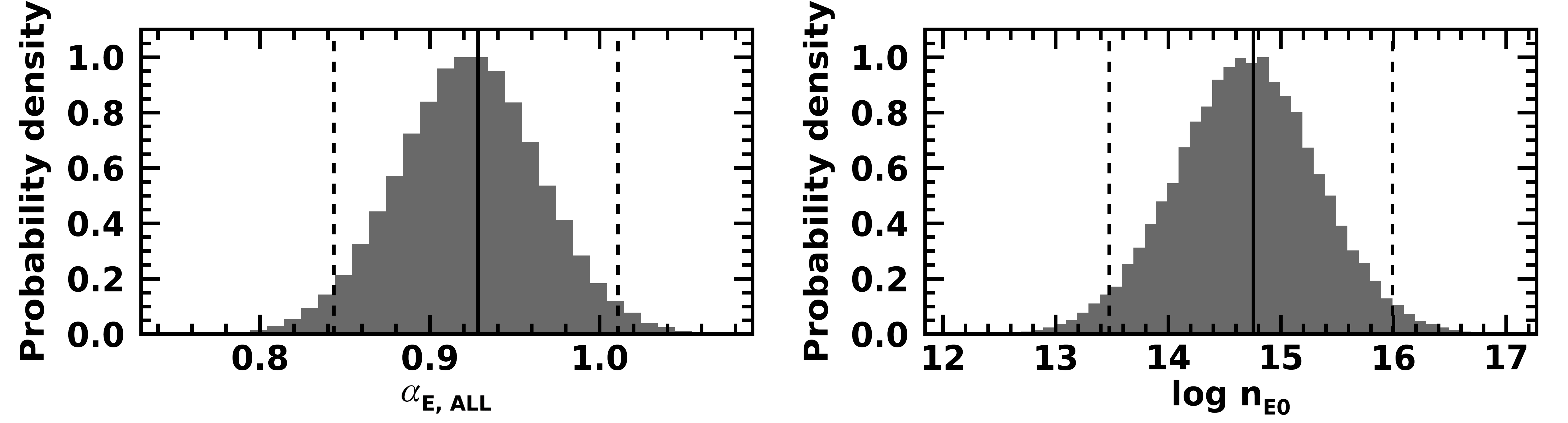}
\caption{Marginal posterior distributions of the power law index and constant of the best fitting shown in Figures \ref{fig:frequency} and \ref{fig:energy}. Each panel corresponds to the energy flux-frequency distribution (top), the number-energy flux distribution (middle), and the number-energy distribution (bottom). Solid lines show a maximum a posterior probability that we use as the slope results and dashed lines show 95\% credible intervals.}
\label{fig:posterior}
\end{figure}

We demonstrate the power law fits of the energy flux-frequency, $s(\omega)$, and number-energy properties distributions, $n(F)$ and $n(E)$. The distributions are fitted with linear functions in log-log scales using an IDL routine mcmc\_fit.pro from the SOBAT\footnote{https://github.com/Sergey-Anfinogentov/SoBAT}. The likelihood function used for the analysis is Gaussians \citep{2021ApJS..252...11A}. We use the normal distribution as the prior for the slope and the specific keyword is as follows: prior\_normal(-2d, 2d) for all three distributions. The uniform distribution is considered as the prior for the fitting constant and the specific keywords are as follows: prior\_uniform(-10d, 10d) for $s(\omega)$, prior\_uniform(-18d, 18d) for $n(F)$, and prior\_uniform(-5d, 5d) for $n(E)$. The best fitting coefficients and their credible intervals are given by a maximum a posterior probability and 95\% credible intervals (Figure \ref{fig:posterior}). The number of samples to generate using Markov Chain Monte Carlo is 100000 and the number of burn-in samples, which is needed for the sampler to find the high probability region and to optimize sampling parameters, is 10000. 

\section{The Influence of Bin Size on the Power Law Slope}

 We investigate the influence of bin size of the distribution on the power law slope results. As shown in Table \ref{tab:binsize}, the slopes depend on the bin size but the difference is insignificant. Consistent slope results that all $\delta$ is less than the critical slope of 1 and all $\alpha_{F}$ and $\alpha_{E}$ is less than the critical slope of 2 for all bin sizes suggest the robustness of our results.
 
\begin{table}[]
\centering
\caption{The influence of the bin size on the power law slopes. The bin size is logarithmic. $\delta_{\text{ALL}}$ is a power law slope of energy flux-frequency distribution for all oscillation events, $\alpha_{F, \text{ALL}}$ number-energy flux distribution, and $\alpha_{E, \text{ALL}}$ number-energy distribution. $\mu$ and $\sigma$ is the average and standard deviation of the values for all bin size.}
\label{tab:binsize}
\begin{tabular}{cccc}
\hline
Bin size         & $\delta_{\text{ALL}}$     & $\alpha_{F, \text{ALL}}$  & $\alpha_{E,\text{ALL}}$  \\ \hline
0.05             & -1.44        & 1.00        & 0.96        \\
0.10             & -1.40        & 1.00        & 0.93        \\
0.20             & -1.40        & 1.02        & 0.97        \\ \hline
$\mu$ ($\sigma$) & -1.41 (0.02) & 1.01 (0.01) & 0.95 (0.02) \\ \hline
\end{tabular}
\end{table}

\bibliography{Lim_bib}{}
\bibliographystyle{aasjournal}

\end{document}